\documentclass[10pt,conference]{IEEEtran}
\IEEEoverridecommandlockouts
\usepackage{cite}
\usepackage[hidelinks]{hyperref}

\usepackage{amsmath,amssymb,amsfonts}
\usepackage{amsthm}
\usepackage{algorithmic}

\usepackage[table]{xcolor}
\usepackage{colortbl}

\usepackage{graphicx}
\usepackage{textcomp}
\usepackage{booktabs}
\usepackage{makecell}
\usepackage{multirow}
\usepackage{array}
\usepackage{tabularx}
\usepackage{adjustbox}
\usepackage{pifont}
\usepackage{subcaption}
\usepackage{arydshln}
\usepackage{xspace}
\usepackage{ragged2e}
\usepackage{tikz}

\usepackage{booktabs}
\usepackage{multirow}
\usepackage{makecell}
\usepackage{threeparttable}
\usepackage[most]{tcolorbox}
\usepackage{enumitem}

\definecolor{FailurePurple}{HTML}{F9DDCA}

\definecolor{CostPeach}{HTML}{F9DDCA}
\definecolor{TokenMint}{HTML}{BFEFEA}
\definecolor{HeaderGray}{HTML}{F3F0EC}

\definecolor{GPTFourBg}{HTML}{F7F7F7} 
\definecolor{GPTFiveBg}{HTML}{FFFFFF} 
\definecolor{DeepSeekBg}{HTML}{F7F7F7}  
\definecolor{MiniMaxBg}{HTML}{FFFFFF} 

\newcommand{\circled}[1]{\textcircled{\scriptsize #1}}

\newcommand{\modelbg}[2]{\cellcolor{#1!100}#2}

\newtheorem{definition}{Definition}

\newcommand{\dataset}{\textit{LiLCVE}\xspace}

\newcommand{\cmark}{\checkmark}
\newcommand{\xmark}{$\times$}

\newcommand{\dopen}{\tikz[baseline=-0.55ex,x=0.75ex,y=0.75ex]{%
  \draw[line width=0.45pt] (0,1)--(1,0)--(0,-1)--(-1,0)--cycle;}}
\newcommand{\dhalf}{\tikz[baseline=-0.55ex,x=0.75ex,y=0.75ex]{%
  \fill (0,1)--(1,0)--(0,-1)--cycle;
  \draw[line width=0.45pt] (0,1)--(1,0)--(0,-1)--(-1,0)--cycle;}}
\newcommand{\dfull}{\tikz[baseline=-0.55ex,x=0.75ex,y=0.75ex]{%
  \fill (0,1)--(1,0)--(0,-1)--(-1,0)--cycle;}}

\newcommand{\copen}{\tikz[baseline=-0.55ex,x=0.75ex,y=0.75ex]{%
  \draw[line width=0.45pt] (0,0) circle (0.9);}}
\newcommand{\chalf}{\tikz[baseline=-0.55ex,x=0.75ex,y=0.75ex]{%
  \fill (0,0)--(90:0.9) arc (90:-90:0.9)--cycle;
  \draw[line width=0.45pt] (0,0) circle (0.9);}}
\newcommand{\cfull}{\tikz[baseline=-0.55ex,x=0.75ex,y=0.75ex]{%
  \fill (0,0) circle (0.9);}}

\newcolumntype{Y}{>{\RaggedRight\arraybackslash}X}

\newcommand{\forestcell}[3]{%
\begin{tikzpicture}[
  x=0.90cm,
  y=0.18cm,
  baseline=-0.6ex
]
  \path[use as bounding box]
    (0,-0.8) rectangle ({ln(8)},0.8);

  \draw[gray!30,line width=0.3pt]
    (0,-0.6) -- (0,0.6);
  \draw[gray!30,line width=0.3pt]
    ({ln(2)},-0.6) -- ({ln(2)},0.6);
  \draw[gray!70,dashed,line width=0.5pt]
    ({ln(4)},-0.75) -- ({ln(4)},0.75);
  \draw[gray!30,line width=0.3pt]
    ({ln(8)},-0.6) -- ({ln(8)},0.6);

  \draw[line width=0.65pt]
    ({ln(#2/0.25)},0) -- ({ln(#3/0.25)},0);
  \draw[line width=0.65pt]
    ({ln(#2/0.25)},-0.22) -- ({ln(#2/0.25)},0.22);
  \draw[line width=0.65pt]
    ({ln(#3/0.25)},-0.22) -- ({ln(#3/0.25)},0.22);

  \fill ({ln(#1/0.25)},0) circle (1.5pt);
\end{tikzpicture}%
}

\newcommand{\up}[1]{\textcolor[RGB]{206, 128, 106}{$\uparrow$#1}}
\newcommand{\down}[1]{\textcolor[RGB]{89, 184, 180}{$\downarrow$#1}}

\newcommand{\val}[1]{#1}
\newcommand{\cellup}[2]{\makecell[r]{#1\\[-1pt]\scriptsize\up{#2}}}
\newcommand{\celldown}[2]{\makecell[r]{#1\\[-1pt]\scriptsize\down{#2}}}

\def\BibTeX{{\rm B\kern-.05em{\sc i\kern-.025em b}\kern-.08em
    T\kern-.1667em\lower.7ex\hbox{E}\kern-.125emX}}

\begin{document}

\title{When LLMs Join the Exploit Chain: Towards Demystifying and Repairing LLM-in-the-Loop Vulnerabilities}

\title{Towards Demystifying and Repairing LLM-in-the-Loop Vulnerabilities}

\author{
\IEEEauthorblockN{Yujie Ma\IEEEauthorrefmark{1},
Jialin Rong\IEEEauthorrefmark{1},
Chenxi Yang\IEEEauthorrefmark{1},
Lili Quan\IEEEauthorrefmark{2},
Jin Wen\IEEEauthorrefmark{3},
Xiaofei Xie\IEEEauthorrefmark{2},
Yongqiang Lyu\IEEEauthorrefmark{1},
Qiang Hu\IEEEauthorrefmark{1}
}
\IEEEauthorblockA{\IEEEauthorrefmark{1}Tianjin University }
\IEEEauthorblockA{\IEEEauthorrefmark{2}Singapore Management University}
\IEEEauthorblockA{\IEEEauthorrefmark{3}University of Luxembourg}
}


\maketitle

\begin{abstract}

Large Language Models~(LLMs) have been actively integrated into modern software systems as critical components, introducing a new type of software vulnerability, LLM-in-the-Loop~(LiL) vulnerability, in which threats are caused by LLMs. Although some studies have attempted to investigate the impact of LiL vulnerabilities, they have unfortunately failed to clearly distinguish LiL vulnerabilities from conventional ones, leaving the understanding of real-world LiL vulnerabilities an open problem.

To address this gap, we first clearly define the scope of LiL vulnerability, and discuss the differences between LiL vulnerabilities and vulnerabilities that exist in LLM systems but are not really caused by LLMs~(i.e., LLM-ecosystem vulnerabilities). Then, we construct the first LiL vulnerability dataset, \dataset, covering 41 LiL vulnerabilities and 75 LLM-ecosystem vulnerabilities, to facilitate the risk analysis of LLM-integrated software. The analysis of \dataset reveals that LiL vulnerabilities have higher severity than LLM-ecosystem vulnerabilities and conventional software vulnerabilities, with 15.5\% and 30.3\% more critical vulnerabilities, respectively.

Furthermore, given the high severity of LiL vulnerabilities and the potential of LLM-based vulnerability repair methods in patching conventional software vulnerabilities. We explore the capabilities of existing widely-used LLM-based methods in repairing vulnerabilities in \dataset. Experimental results on 20 agent–model configuration demonstrate that LiL vulnerabilities are far more challenging to fix, with an average decrease of 10.8\% Pass@1 rate compared to other types of vulnerabilities. More critically, three categories, Generated Query Execution, Agent Action, and Model Output Rendering, frequently receive 0\% repair success rates. 




\end{abstract}

\begin{IEEEkeywords}
Large Language Model, Vulnerability Analysis, Vulnerability Repair
\end{IEEEkeywords}

\section{Introduction}

Large Language Models~(LLMs) have achieved remarkable progress, demonstrating strong capabilities across diverse domains and increasingly transforming modern software engineering practices~\cite{dozono2026llmforassessment,jiang2026surveycodegen,wang2024softwaresurvey}. Unfortunately, LLMs are also known to exhibit robustness limitations, such as prediction uncertainty and hallucinations, which can lead to incorrect or harmful outputs and ultimately compromise the reliability of the integrated systems. For example, a code injection flaw in Langroid's
\texttt{LanceDocChatAgent} allows an attacker to induce the agent to place a malicious expression in \texttt{QueryPlan.dataframe\_calc}, which is evaluated by \texttt{pandas.eval()} and may result in arbitrary command execution and host compromise~(CVE-2025-46725~\cite{CVE-2025-46725}).

Motivated by these risks, recent studies have attempted to analyze vulnerabilities in LLM supply chains, LLM applications, agent systems, and supporting components across different stages of the LLM development lifecycle~\cite{wang2025sok, ma2025understanding, shen2025security}. However, due to the unclear definition of LLM-relevant vulnerabilities, most seemingly LLM-relevant vulnerabilities are in fact conventional software weaknesses, such as path traversal, unsafe deserialization, and authentication errors~\cite{mitreCWE22, mitreCWE918, mitreCWETop25}. That is, even though these flaws appear in LLM-related projects, their exploitation does not necessarily require model invocation or model-mediated behavior. 
For example, in LLMSCBench~\cite{ma2025understanding}, only 11 out of 1,229 vulnerabilities involve LLMs as a necessary step in the exploitation chain. As a result, vulnerabilities genuinely caused by LLMs have not been deeply analyzed. 

To bridge this gap, in the paper, we first define \textit{LLM-in-the-loop~(LiL) vulnerability}, which is a software vulnerability whose exploitation must involve an LLM, and \textit{LLM-ecosystem vulnerability}, which is in LLM-integrated systems but not exactly triggered by LLMs. Based on this definition, we construct the first LiL dataset, \dataset, to facilitate the understanding of LiL and LLM-ecosystem vulnerabilities. Concretely, we collect 2,888 vulnerabilities associated with 230 open-source LLM components from huntr~\cite{huntr}. Five researchers manually review their descriptions, PoCs, issue discussions, and patches, identifying 116 vulnerabilities, including 41 LiL vulnerabilities and 75 LLM-ecosystem vulnerabilities.  Based on the analysis of \dataset, we aim to answer the following research question:

\begin{itemize}[leftmargin=*]
    \item \textbf{RQ1(Characteristic):} \textit{What are the differences between LiL vulnerabilities, LLM-ecosystem vulnerabilities, and conventional vulnerabilities?} We analyze the CWE type and the severity of each set of vulnerabilities to figure out their fundamental difference.

\end{itemize}

The analysis results reveal that LiL vulnerabilities differ from the other two types of vulnerabilities in both type distribution and severity.
LLM-ecosystem and conventional software vulnerabilities are mainly dominated by conventional weaknesses such as CWE-22 (Path Traversal), CWE-79 (Cross-Site Scripting) and CWE-400 (Uncontrolled Resource Consumption). In contrast, LiL vulnerabilities are more concentrated in injection- and execution-related weaknesses such as CWE-94(Code Injection) and CWE-74(Injection), reflecting model-mediated risks at downstream trust boundaries. Moreover, LiL vulnerabilities exhibit higher severity than LLM-ecosystem and conventional ones, with critical cases accounting for 51.2\%.

Given the high severity of LiL vulnerabilities and the goal of vulnerability analysis to improve software reliability, we investigate the capabilities of existing widely used LLM-based vulnerability repair methods on \dataset. To do so, we enrich \dataset to enable the repair evaluation, e.g., preparing test cases for patch validation. Based on \dataset, we evaluate five repair agents~(e.g., SWE-Agent~\cite{yang2024swe}) with four foundation models~(e.g., GPT-5.2-Codex~\cite{gpt52codex}), resulting in 20 agent--model configurations, and aim to answer the following research questions.

\begin{itemize}[leftmargin=*]
    \item \textbf{RQ2 (Effectiveness):} \textit{What is the effectiveness of LLM-based vulnerability repair methods on \dataset?} We compare the repair correctness~(Pass@1) of agents across LiL, LLM-ecosystem, and conventional vulnerabilities followed by regression and patch-complexity analysis.

    \item \textbf{RQ3 (Efficiency):} \textit{What is the cost of fixing vulnerabilities in \dataset?}  We measure the monetary cost and the number of repair steps required to assess the efficiency of agents in repairing vulnerabilities in \dataset.

    \item \textbf{RQ4 (Root Cause):} \textit{What is the root cause that leads to the failed vulnerability patching?} We analyze the failed repair outcomes and representative patches to investigate the dominant mechanisms underlying repair failures.
\end{itemize}


Our results show that \dataset vulnerabilities are substantially more challenging for existing automated repair agents to fix. Across all 20 agent--model configurations, \dataset vulnerabilities achieve lower Pass@1 than the conventional control vulnerabilities, with an average drop of 10.8\%. The difficulty also varies across vulnerability categories. Generated Query Execution, Agent or Tool Action, and Model Output Rendering are particularly more challenging, with 16 out of 20 configurations achieving a 0\% Pass@1 in at least one category. Moreover, repairing LiL vulnerabilities requires greater effort in terms of repair steps. Finally, the failure analysis reveals that on average, 72.3\% of the generated patches can only pass security or functional testing. Multi-objective satisfaction is a key challenge in fixing LiL vulnerabilities.

In summary, this paper makes the following contributions:

\begin{itemize}[leftmargin=*]
    \item We are the first to define LLM-in-the-loop~(LiL) vulnerabilities by modeling the role of LLMs in vulnerability exploitation, thereby distinguishing them from both LLM-ecosystem vulnerabilities without LLM involvement and conventional software vulnerabilities.

    \item We construct \dataset, the first executable benchmark datasets with 41 LiL vulnerabilities and 75 LLM-ecosystem vulnerabilities, enabling the automatic evaluation of vulnerability repair capabilities of LLM-based methods. 

    \item We comprehensively evaluate 20 agent--model configurations using \dataset and reveal that LiL vulnerabilities are consistently more challenging and costly to repair than LLM-ecosystem vulnerabilities.

    \item Our failure analysis shows that existing agents often fail to correctly mediate the boundary between model outputs and downstream system behaviors, providing clear directions for the future design of automated repair tools tailored to LLM supply chain security.
\end{itemize}

\begin{figure*}[!t]
    \centering
    \includegraphics[width=0.98\textwidth]{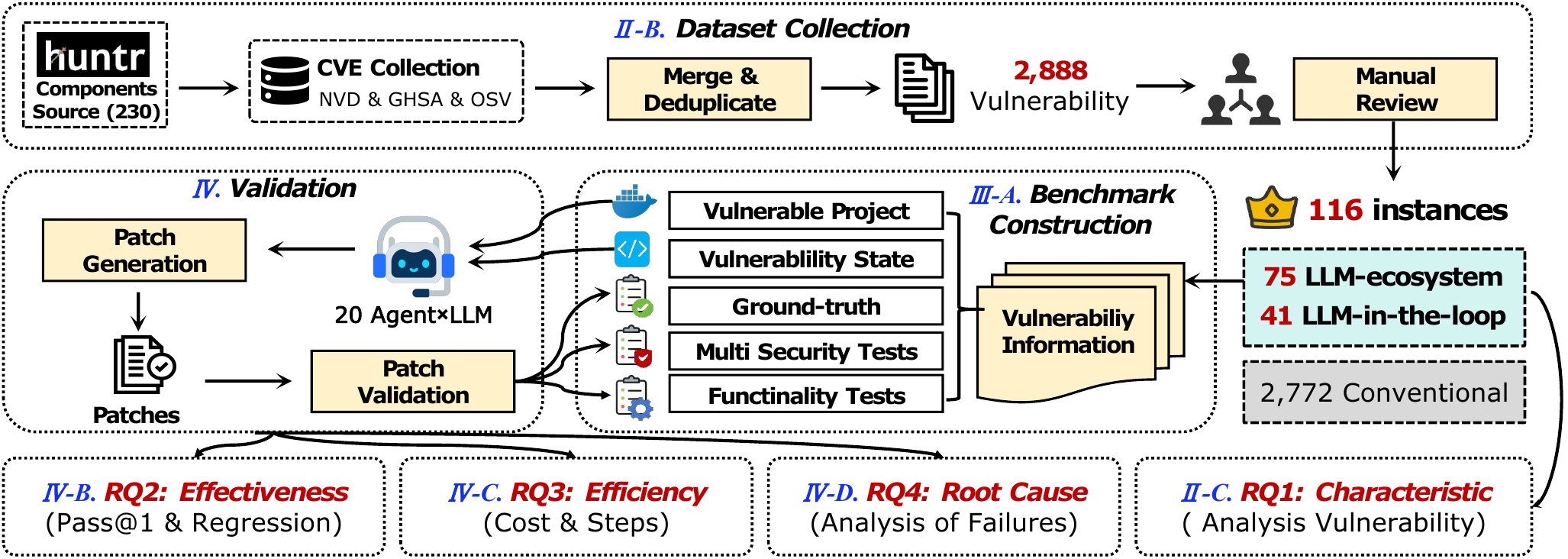}  
    \caption{Overview of our study workflow.}
    \label{fig:workflow}
\end{figure*}

\section{LiL Vulnerability Study}

\subsection{Definition}
\label{sec:definition-of-llmloop}


There are limited definitions for LLM-relevant vulnerabilities, although Shen et al.~\cite{shen2025security} define \emph{LLM2Tool vulnerabilities}, their definition is limited to flows from LLM outputs to tools. Other definitions tailored to particular exploitation patterns~\cite{AgentFuzz, taintp2x, LLMSmith}, but primarily serve their respective detection tasks and are not intended as a unified criterion for constructing a general vulnerability dataset. 

As a result, existing LLM-relevant vulnerability datasets are often constructed based on affected projects or components, rather than whether LLMs participate in exploitation. They therefore mix vulnerabilities triggered by LLM action with many conventional software flaws, and cannot isolate the distinct vulnerability distribution
and threat surface that emerges when LLMs participate in exploitation.  For example, CVE-2024-11394~\cite{CVE-2024-11394} in LLMSCBench\cite{ma2025understanding} affects model-file handling in Hugging Face Transformers, a widely used library for LLMs. A crafted Trax model file can trigger arbitrary code execution
through unsafe deserialization. However, the exploit occurs while loading the model and requires neither LLM inference nor model output. Its root cause is conventional deserialization of untrusted
data (CWE-502), a software weakness extensively studied before the emergence of LLMs.

To address this gap, we introduce definitions of LLM-in-the-Loop vulnerability, and LLM-ecosystem vulnerabilities.

\begin{definition}[\textbf{LLM-in-the-Loop~(LiL) Vulnerability}] 
\label{def:lil-vulnerability}
A software vulnerability triggered by the action of LLMs.
\end{definition}


The \textit{action} denotes how an LLM participates in the exploit chain of a software vulnerability. In traditional software vulnerability analysis, an exploit chain typically starts from attacker-controlled input, passes through intermediate program logic, and eventually reaches a security-sensitive sink~\cite{livshits2005finding}. When an LLM becomes part of this chain, its influence can be characterized according to where and how it changes this input--logic--sink process. Specifically, the \textit{action} is threefold \circled{1} \emph{LLM-generated artifacts}. Prompt injection or model-facing input causes the LLM to generate code, expressions, SQL/Cypher, shell commands, or other executable or interpretable payloads that reach a downstream sink, e.g., CVE-2023-29374~\cite{CVE-2023-29374} in LangChain \texttt{LLMMathChain}. \circled{2} \emph{LLM-mediated decisions or actions}. Exploitation depends on the LLM selecting tools, constructing tool arguments, routing execution, or triggering privileged agent actions, e.g., CVE-2025-5273~\cite{CVE-2025-5273} in \texttt{mcp-markdownify-server}. \circled{3} \emph{LLM-output consumption}. Model output is rendered, parsed, dispatched, or evaluated by downstream logic in a security-sensitive way, e.g., CVE-2025-9141~\cite{CVE-2025-9141} in vLLM's tool-call parser.



\begin{definition}[\textbf{LLM-ecosystem Vulnerability}] A software vulnerability in LLM-integrated systems, but not triggered by the action of LLMs.
\end{definition}

Simply speaking, LLM-ecosystem vulnerabilities are those in the LLM-relevant ecosystems but not LiL vulnerabilities. 

\textbf{Representative LLM-ecosystem vulnerabilities:}
\circled{1} path traversal or arbitrary file-write bugs reachable through ordinary API inputs, e.g., CVE-2024-7034~\cite{CVE-2024-7034} in Open WebUI; 
\circled{2} SSRF or network-request bugs controlled directly by parameters or configuration fields, e.g., CVE-2024-6587~\cite{CVE-2024-6587} in LiteLLM; 
\circled{3} model-file, checkpoint, or dependency deserialization vulnerabilities exploitable by loading untrusted artifacts, e.g., CVE-2024-12029~\cite{CVE-2024-12029} in InvokeAI; and 
\circled{4} ReDoS or resource-exhaustion bugs in ordinary parsers, converters, or utility functions, e.g., CVE-2025-5197~\cite{CVE-2025-5197} in Hugging Face Transformers.

\subsection{Dataset Construction}

\subsubsection{Data Collection}

First, we need to identify the target environments where LiL vulnerabilities and LLM-ecosystem vulnerabilities might occur. To do so, we choose targets provided by Huntr~\cite{huntr}, the world's biggest bug bounty platform dedicated to AI/ML. Due to the high bounties, these vulnerabilities have critical value.
In total, in the first stage, we extract all~(230) open-source components~(targets) listed in Huntr.

Then we use extracted component names as matching keys and collect vulnerabilities from three major vulnerability databases: National Vulnerability Database~(NVD), GitHub Advisory Database~(GHSA), and Open Source Vulnerabilities~(OSV). To reduce noisy matches, we match the keywords against structured affected-product fields, such as CPE entries in NVD and affected package fields in GHSA and OSV. Through this step, we retrieve 1,489 vulnerabilities from the GHSA, 550 from the NVD, and 1,751 from the OSV database. To remove duplicates, we first merge records using explicit alias fields~(e.g., mapping GHSA IDs to CVE IDs). Then, we cross-reference shared metadata~(e.g., identical reference URLs pointing to the same GitHub pull requests, issues, or commit hashes) to eliminate any remaining redundancies. For records lacking explicit cross-references, we manually review the relevant information. This process yields an initial set of 2,888 unique vulnerabilities. Each record contains essential metadata, including summary, detailed description, CWE, severity, CVSS score, CPE, references, patch links, affected versions, and time information.

\begin{table*}[t]
\centering
\caption{Comparison with Existing Vulnerability and Repair Evaluation Datasets. ST = security test; FT = functionality test.
$\protect\dopen$, $\protect\dhalf$, and $\protect\dfull$ denote no, coarse filtering by metadata such as year and component, and verified LLM necessity, respectively.
$\protect\copen$, $\protect\chalf$, and $\protect\cfull$ denote no test, tests without multiple variants, and tests with multiple variants, respectively.
Numbers in parentheses indicate the total number of corresponding test cases. * indicates vulnerabilities with associated tests, but the exact test counts are unavailable. NA indicates that the corresponding information could not be identified from available sources.}
\label{tab:dataset-comparison}
\scriptsize
\setlength{\tabcolsep}{2.0pt}
\renewcommand{\arraystretch}{1.12}

\begin{tabular*}{\textwidth}{@{\extracolsep{\fill}}
>{\centering\arraybackslash}m{0.116\textwidth}
>{\raggedright\arraybackslash}m{0.111\textwidth}
>{\centering\arraybackslash}m{0.061\textwidth}
>{\centering\arraybackslash}m{0.061\textwidth}
>{\centering\arraybackslash}m{0.045\textwidth}
>{\centering\arraybackslash}m{0.049\textwidth}
>{\centering\arraybackslash}m{0.074\textwidth}
>{\centering\arraybackslash}m{0.069\textwidth}
>{\centering\arraybackslash}m{0.074\textwidth}
>{\centering\arraybackslash}m{0.069\textwidth}
>{\centering\arraybackslash}m{0.069\textwidth}
@{}}
\toprule[1.2pt]
\multirow{2}{*}{\textbf{Group}}
& \multirow{2}{*}{\textbf{Dataset}}
& \multirow{2}{*}{\textbf{\#Component}}
& \multirow{2}{*}{\textbf{\#Vuln.}}
& \multirow{2}{*}{\textbf{\#CWE}}
& \multirow{2}{*}{\textbf{\#Patch}}
& \multirow{2}{*}{\makecell[c]{\textbf{Vuln.}\\\textbf{Period}}}
& \multirow{2}{*}{\makecell[c]{\textbf{LLM-}\\\textbf{ecosystem?}}}
& \multirow{2}{*}{\makecell[c]{\textbf{LLM}\\\textbf{necessity?}}}
& \multicolumn{2}{c}{\textbf{Repair Evaluation?}} \\
\cmidrule(lr){10-11}
& & & & & & & & &
\textbf{multi ST}
& \textbf{multi FT} \\
\midrule
\multirow{4}{=}{\makecell[c]{LLM-ecosystem\\security studies}}
& Wang et al.~\cite{wang2025sok} & 75 & 529 & 28 & 300 & 2023--2024 & \cmark & \dhalf & \copen & \copen \\
& Hu et al.~\cite{hu2025understanding} & NA & 180 & NA & NA & 2023--2025 & \cmark & \dhalf & \copen & \copen \\
& Ma et al.~\cite{ma2025understanding} & 3,373 & 1,229 & 145 & 1152 & 2008--2025 & \cmark & \dhalf & \copen & \copen \\
& Shen et al.~\cite{shen2025security} & 50 & 221 & 60 & 159 & 2023--2025 & \cmark & \dhalf & \copen & \copen \\

\midrule
\multirow{4}{=}{\makecell[c]{Traditional vul.\\repair benchmarks\\(for LLM/Agent)}}
& CVE-Bench~\cite{wang2025cve} & 120 & 509 & 105 & 509 & 2008--2022 & \xmark & \dopen & \chalf~(509*) & \copen \\
& VulnRepairEval~\cite{wang2025vulnrepaireval} & NA & 23 & 16 & 23 & 2017--2024 & \xmark & \dopen & \chalf~(23*) & \copen \\
& PatchEval~\cite{wei2025patcheval} & NA & 1,000 & 65 & 1000 & 2015--2025 & \xmark & \dopen & \chalf~(230) & \chalf~(176*) \\
& SEC-bench~\cite{lee2026sec} & 29 & 200 & 17 & 200 & 2016--2024 & \xmark & \dopen & \chalf~(200) & \copen \\

\midrule

\multirow{3}{=}{\makecell[c]{Traditional vul.\\repair benchmarks\\(general)}}
& ARVO~\cite{mei2024arvo} & 273 & 5,001 & NA & 5001 & 2016--2024 & \xmark & \dopen & \chalf~(5,001*) & \copen \\
& Vul4J~\cite{bui2022vul4j} & 51 & 79 & 25 & 79 & 2012--2021 & \xmark & \dopen & \chalf~(126) & \chalf~(79*) \\
& Vul4C~\cite{hu2025sokVul4C} & 23 & 144 & 19 & 144 & 2010--2023 & \xmark & \dopen & \chalf~(144) & \chalf~(68*) \\

\midrule

\makecell[c]{LLM vul.\\ repair benchmark\\(for LLM/Agent)}
& \textbf{\dataset (ours)}
& \textbf{230}
& \textbf{2,888}
& \textbf{209}
& \textbf{116}
& \textbf{2023--2025}
& \textbf{\cmark}
& \textbf{\dfull}
& \textbf{\cfull~(429)}
& \textbf{\cfull~(411)} \\

\bottomrule[1.2pt]
\end{tabular*}

\end{table*}


\begin{figure*}[t]
  \centering
  \includegraphics[width=1\textwidth]{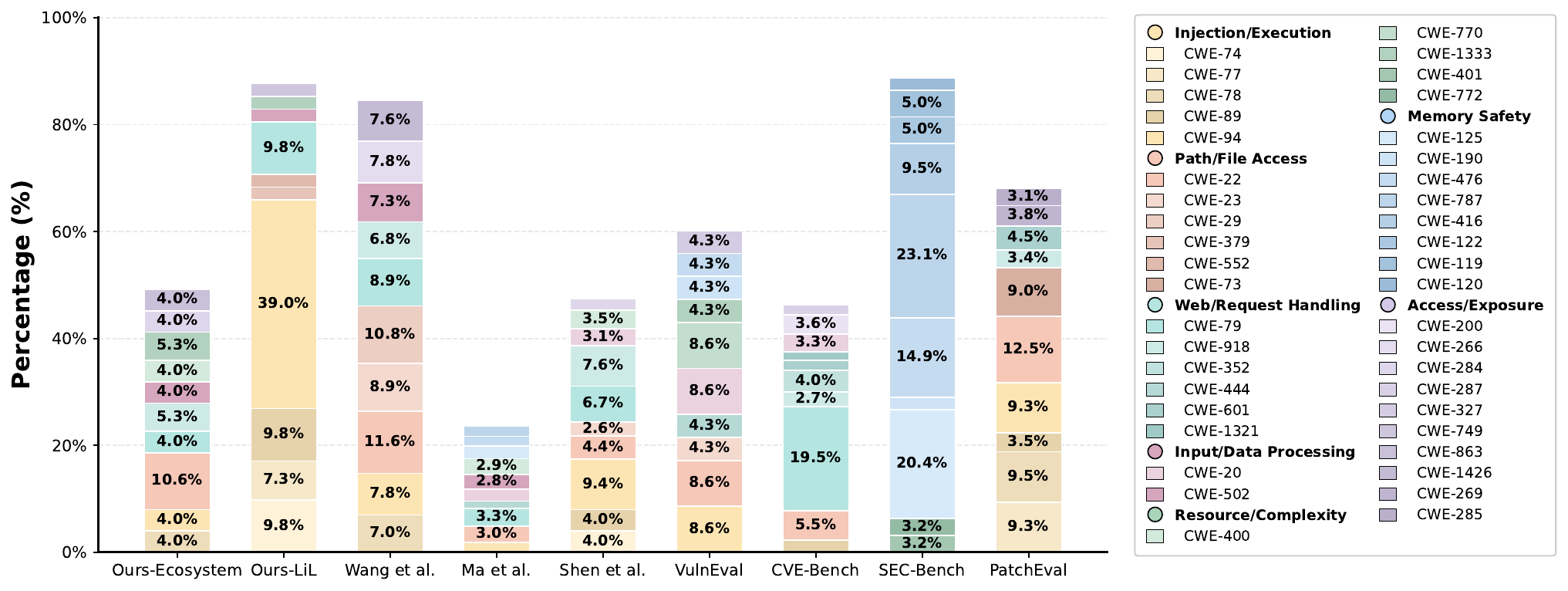}
   \caption{Top CWE distributions across datasets. Percentages below 2.5\% are omitted from labels for readability.}
  \label{fig:cwe-distribution}
\end{figure*}

\begin{figure*}[t]
  \centering

  \begin{subfigure}{0.20\textwidth}
    \centering
    \includegraphics[width=\linewidth]{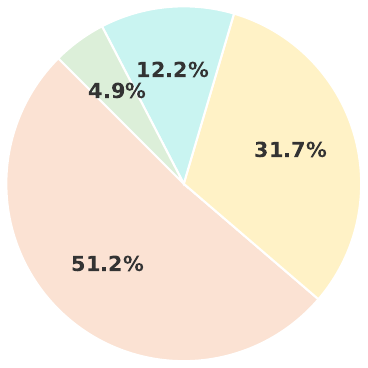}
    \caption{\dataset-LiL}
  \end{subfigure}
  \hfill
  \begin{subfigure}{0.20\textwidth}
    \centering
    \includegraphics[width=\linewidth]{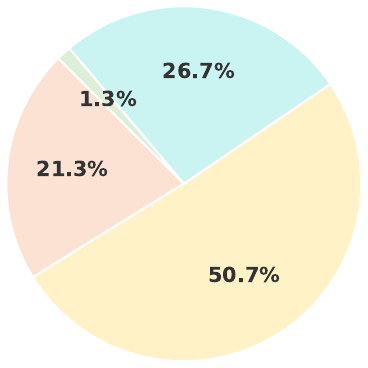}
    \caption{\dataset-ecosystem}
  \end{subfigure}
  \hfill
  \begin{subfigure}{0.20\textwidth}
    \centering
    \includegraphics[width=\linewidth]{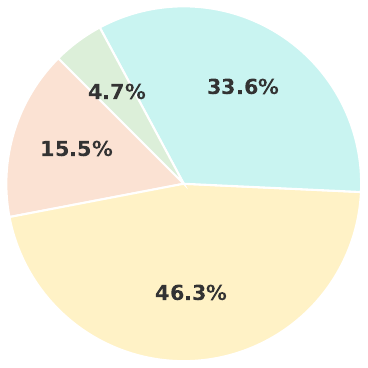}
    \caption{Ma et al.}
  \end{subfigure}
  \hfill
  \begin{subfigure}{0.20\textwidth}
    \centering
    \includegraphics[width=\linewidth]{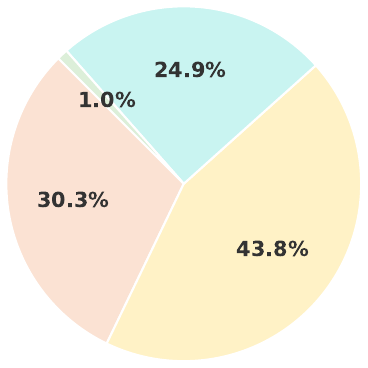}
    \caption{Shen et al.}
  \end{subfigure}
  \vspace{0.5em}   
  \begin{subfigure}{0.20\textwidth}
    \centering
    \includegraphics[width=\linewidth]{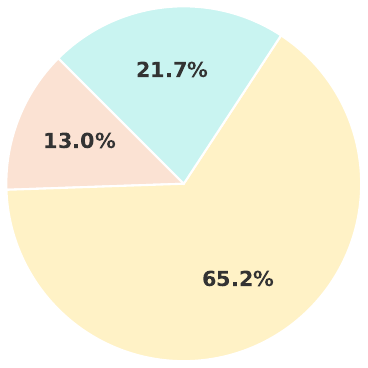}
    \caption{VulnEval}
  \end{subfigure}
  \hfill
  \begin{subfigure}{0.20\textwidth}
    \centering
    \includegraphics[width=\linewidth]{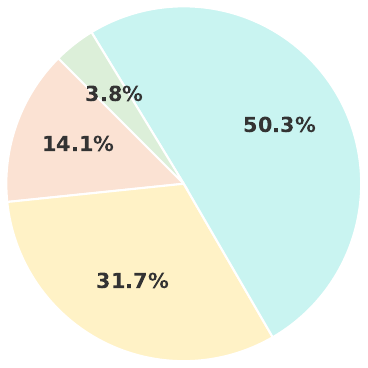}
    \caption{CVE-Bench}
  \end{subfigure}
  \hfill
  \begin{subfigure}{0.20\textwidth}
    \centering
    \includegraphics[width=\linewidth]{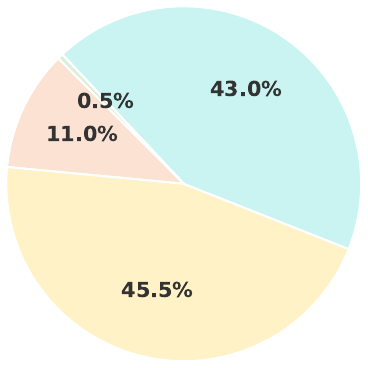}
    \caption{SEC-Bench}
  \end{subfigure}
  \hfill
  \begin{subfigure}{0.20\textwidth}
    \centering
    \includegraphics[width=\linewidth]{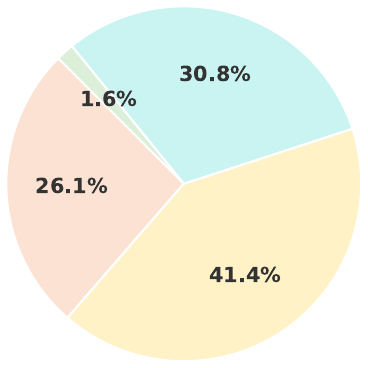}
    \caption{PatchEval}
  \end{subfigure}
  \vspace{0.4em}

  \includegraphics[width=0.36\textwidth]{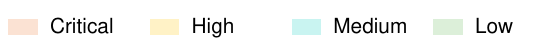}

  \caption{Risk severity distributions across datasets. \dataset-LiL and \dataset-ecosystem indicate LiL and LLM-ecosystem vulnerabilities in \dataset.}
  \label{fig:severity-distribution}
\end{figure*}

\subsubsection{Data Selection}

We then follow our definition~(Section~\ref{sec:definition-of-llmloop}) to identify the target vulnerabilities. This step cannot be reliably automated, as existing vulnerability records cannot precisely distinguish whether vulnerabilities are exactly triggered by LLMs or not. Therefore, five security researchers, each with more than three years of experience in vulnerability analysis, carries out a three-step manual analysis process: (1) identifying the attacker-controlled entry point and determining whether the vulnerability is related to LLMs (to identify LLM-ecosystem vulnerability); (2) reconstructing the exploit chain by tracing how attacker-controlled inputs, model outputs, tool calls, and application logic interact before reaching the vulnerable operation (to identify potential LiL vulnerability); and 
(3) assessing whether a vulnerability is a LiL vulnerability according to our definition~\ref{def:lil-vulnerability}.
Cases with insufficient evidence are conservatively excluded from the LiL vulnerability set. To assess annotation reliability, each vulnerability was independently labeled by five researchers before discussion. Disagreements were resolved through group discussion until a consensus was reached.

As a result, this audit shows that the majority of collected vulnerabilities are not LiL vulnerabilities. Among the 2,888 raw vulnerabilities, 2,772 are unrelated to LLMs. Only 116 are related to the LLM-ecosystem vulnerability, of which 41 unique vulnerabilities satisfy our LiL definition.


\subsection{RQ1: Characteristics of LiL Vulnerabilities}


\textbf{CWE Type.} Figure~\ref{fig:cwe-distribution} shows that LLM-ecosystem vulnerabilities share a similar CWE distribution with conventional vulnerabilities, especially for path and file access (CWE-22), web request handling (CWE-79 and CWE-918), and resource exhaustion (CWE-400). This similarity indicates that most of the LLM ecosystem vulnerabilities remain conventional program-level defects. In contrast, LiL vulnerabilities exhibit a markedly concentrated distribution, with injection and execution weaknesses (CWE-94, CWE-89, CWE-74, and CWE-77) accounting for 65.9\%. This concentration highlights a distinct vulnerability pattern introduced by placing LLMs inside the execution or decision loop, where model-generated or model-interpreted content can affect downstream application logic.




\textbf{Taxonomy based on LiL definition.} CWE categories cannot fully capture how the LLM participates in the vulnerable execution chain. To better characterize these LLM-specific mechanisms, we further categorize LiL vulnerabilities based on the LiL definition~\ref{def:lil-vulnerability}. Among the 41 LiL vulnerabilities, generated code or expression execution~(C1) is the most common category with 20 cases. Generated query execution~(C2) and agent or tool actions~(C3) each account for 7 cases, followed by model output rendering~(C4) and model output parsing or dispatch~(C5). C1 and C2 correspond to \circled{1} in the LiL definition, C3 corresponds to \circled{2}, and C4 and C5 correspond to \circled{3}. This taxonomy complements the CWE analysis by showing that LiL vulnerabilities are not merely conventional injection bugs in LLM-related projects. Instead, they arise from distinct ways in which LLM outputs or decisions affect downstream components.

\textbf{Severity.} Figure~\ref{fig:severity-distribution} compares severity distributions across three groups, LiL-vulnerabilities~(Subfigure~(a)), LLM-ecosystem vulnerabilities~(Subfigure~(b)--(d)), and conventional vulnerabilities~(Subfigure~(e)--(h)). The results reveal that LLM-ecosystem and conventional vulnerabilities are mainly dominated by High and Medium vulnerabilities. In contrast, \dataset-LiL contains the largest Critical proportion at 51.2\%, which is on average, 28.8\% and 35.1\% more than LLM-ecosystem and conventional software vulnerabilities, respectively. These results indicate the unique pattern and high severity of LiL vulnerabilities, highlighting the additional attention in studying and fixing this type of vulnerabilities.

\begin{tcolorbox}[size=title,opacityfill=0.1,breakable]
\noindent
\textbf{Answer to RQ1:} The CWE distribution of LiL vulnerabilities is significantly different from LLM-ecosystem vulnerabilities and conventional software vulnerabilities. LiL vulnerabilities have, on average, 28.8\% and 35.1\% more Critical instances than LLM-ecosystem and conventional software vulnerabilities.

\end{tcolorbox}

\begin{table}[t]
\centering
\caption{Taxonomy of LiL vulnerabilities.}
\label{tab:llm_vul_taxonomy}
\footnotesize
\setlength{\tabcolsep}{3pt}
\renewcommand{\arraystretch}{1.7}
\begin{tabularx}{\columnwidth}{
    >{\centering\arraybackslash}p{0.28\columnwidth}
    >{\raggedright\arraybackslash}X
    >{\centering\arraybackslash}p{0.10\columnwidth}
}
\toprule[1.2pt]
\textbf{Category} & \textbf{Classification Basis} & \textbf{Count} \\
\midrule
\textbf{C1: Generated Code / Expression Execution}
& LLM-generated code or expressions reach execution sinks such as \texttt{exec}, \texttt{eval}, sandbox, or interpreter.
& 20 \\
\textbf{C2: Generated Query Execution}
& LLM-generated SQL/Cypher/query is executed by a database or query engine.
& 7 \\
\textbf{C3: Agent / Tool Action}
& LLM selects tools or constructs tool arguments for privileged external actions.
& 7 \\
\textbf{C4: Model Output Rendering}
& Model-controlled HTML, Markdown, or UI content is rendered unsafely.
& 4 \\
\textbf{C5: Model Output Parsing / Dispatch}
& Model output is parsed or dispatched by callbacks, tool parsers, or structured-output handlers.
& 3 \\
\midrule
\textbf{Total}
& LLM-in-the-loop vulnerabilities.
& \textbf{41} \\
\bottomrule[1.2pt]
\end{tabularx}
\end{table}

\section{LiL Vulnerability Patching}

As discussed before, most of the LiL vulnerabilities have high risk severity, so fixing them becomes crucial to boost the security of LLM-integrated software systems. Recently, LLM/Agent-based vulnerability repair methods have shown their potential in automatically patching conventional software vulnerabilities. Given this success, we tend to investigate their capabilities in fixing LiL vulnerabilities, hoping to provide some meaningful insights for the future security assurance of LLM-integrated software. 

\subsection{Benchmark Construction}

Automatic vulnerability patching via LLM/Agent requires some necessary inputs beyond basic vulnerability information, such as test cases for correctness validation. Therefore, we enrich \dataset to enable repair evaluation.


\textbf{Version and Patch Identification.}
For each vulnerability, we identify the affected version and fixed version. When a fixing commit contains unrelated changes, we retain only the security-relevant edits as the ground-truth patch. 

\textbf{Vulnerability Statement Construction.} 
For each instance, we curate a \textit{vulnerability statement} to ensure agents receive sufficient and realistic context to understand and repair the vulnerability. This \textit{statement} contains vulnerability information collected from CVE descriptions, security advisories, issue reports, release notes, affected versions, and PoC validation results. To prevent solution leakage, we exclude information related to repairs.

\textbf{Multi-Variant Security Test Construction.}
A PoC often captures one concrete exploit trace and fails to cover all potential exploitation paths. Therefore, for each vulnerability, we first construct a baseline security test from the collected PoC. Then, inspired by prior work on PoC-based bug variant generation~\cite{wei2024sleuth} for precise patch validation, we construct and validate root-cause-preserving exploitation variants. A variant is retained only if it triggers the same vulnerability on the affected version and is blocked by the fixed version. When an exploit involves LLM outputs or agent decisions, we avoid nondeterminism by replaying fixed representative attack payloads. This is reasonable because our tests do not evaluate the frontend jailbreak process, but focus on whether the downstream code logic is correctly fixed.


\textbf{Functionality Test Construction.}
We use functionality tests from the repository that are relevant to the affected functionality and supplement them with targeted tests. A functionality test is retained only if it passes on both the affected version and the fixed version, ensuring that it captures the intended behavior rather than implementation details of the official patch. A generated patch is considered successful only when it builds correctly, blocks all security tests, and passes all functionality tests.

\textbf{Comparison with Existing Datasets and Benchmarks.} Table~\ref{tab:dataset-comparison} compares \dataset with relevant datasets and benchmarks. Existing LLM-ecosystem datasets do not support automatic repair evaluation, while traditional repair benchmarks do not model LLM-dependent exploitation. In contrast, \dataset provides 116 executable repair instances with verified LLM necessity, multi-variant security tests, and functionality tests. 




\section{Vulnerability Repair Evaluation}

\begin{table*}[t]
\centering
\caption{Repair Success Rate~(Pass@1) Comparison Across Traditional Benchmarks, the LLM Ecosystem, and LLM-in-the-Loop Settings. SWE-bench-V and SWE-bench-L denote SWE-bench Verified and SWE-bench Lite, respectively. For the \textit{Traditional Bench} column, results are collected from official benchmark leaderboards and model reports when available. For the \textit{all} column, the arrow reports the absolute change~(\%) relative to Pass@1 under the LLM-ecosystem setting. For c1--c5, the arrow reports the absolute change~(\%) relative to the corresponding \textit{all} score.}
\label{tab:llm_loop_pass1}
\scriptsize
\setlength{\tabcolsep}{2pt}
\renewcommand{\arraystretch}{1.08}
\setlength{\dashlinedash}{2pt}
\setlength{\dashlinegap}{2pt}

\begin{adjustbox}{width=\textwidth}
\begin{tabular}{@{}
>{\raggedright\arraybackslash}p{0.095\textwidth}
>{\raggedright\arraybackslash}p{0.115\textwidth}
>{\raggedleft\arraybackslash}p{0.105\textwidth}
>{\raggedleft\arraybackslash}p{0.095\textwidth}
>{\raggedleft\arraybackslash}p{0.095\textwidth}
>{\raggedleft\arraybackslash}p{0.082\textwidth}
>{\raggedleft\arraybackslash}p{0.082\textwidth}
>{\raggedleft\arraybackslash}p{0.082\textwidth}
>{\raggedleft\arraybackslash}p{0.082\textwidth}
>{\raggedleft\arraybackslash}p{0.082\textwidth}
@{}}
\toprule[1.2pt]
\multirow{2}{*}{\textbf{Agent}}
& \multirow{2}{*}{\textbf{Model}}
& \multirow{2}{*}{\textbf{Traditional Bench}}
& \multirow{2}{*}{\textbf{LLM-ecosystem}}
& \multicolumn{6}{c}{\textbf{LLM-in-the-loop}} \\
\cmidrule(lr){5-10}
& & & & \textbf{all} & \textbf{c1} & \textbf{c2} & \textbf{c3} & \textbf{c4} & \textbf{c5} \\
\midrule

\multirow{4}{*}{OpenHands}
& \modelbg{GPTFourBg}{GPT-4o}        
& \modelbg{GPTFourBg}{\shortstack[r]{22.00\%\\(SWE-bench-L)}}
& \modelbg{GPTFourBg}{\val{17.33\%(13/75)}} 
& \modelbg{GPTFourBg}{\celldown{7.32\%(3/41)}{10.01}}  
& \modelbg{GPTFourBg}{\cellup{10.00\%(2/20)}{2.68}}
& \modelbg{GPTFourBg}{\cellup{14.29\%(1/7)}{6.97}}
& \modelbg{GPTFourBg}{\celldown{0.00\%(0/7)}{7.32}}
& \modelbg{GPTFourBg}{\celldown{0.00\%(0/4)}{7.32}}
& \modelbg{GPTFourBg}{\celldown{0.00\%(0/3)}{7.32}} \\
& \modelbg{GPTFiveBg}{GPT-5.2-Codex}
& \modelbg{GPTFiveBg}{--}
& \modelbg{GPTFiveBg}{\val{56.00\%(42/75)}} 
& \modelbg{GPTFiveBg}{\celldown{53.66\%(22/41)}{2.34}} 
& \modelbg{GPTFiveBg}{\cellup{75.00\%(15/20)}{21.34}} 
& \modelbg{GPTFiveBg}{\celldown{28.57\%(2/7)}{25.09}} 
& \modelbg{GPTFiveBg}{\celldown{28.57\%(2/7)}{25.09}} 
& \modelbg{GPTFiveBg}{\celldown{25.00\%(1/4)}{28.66}} 
& \modelbg{GPTFiveBg}{\cellup{66.67\%(2/3)}{13.01}} \\
& \modelbg{DeepSeekBg}{DeepSeek-V4-Pro}    
& \modelbg{DeepSeekBg}{\shortstack[r]{73.20\%\\(SWE-bench-V)}}
& \modelbg{DeepSeekBg}{\val{60.00\%(45/75)}} 
& \modelbg{DeepSeekBg}{\celldown{41.46\%(17/41)}{18.54}} 
& \modelbg{DeepSeekBg}{\cellup{55.00\%(11/20)}{13.54}} 
& \modelbg{DeepSeekBg}{\celldown{28.57\%(2/7)}{12.89}} 
& \modelbg{DeepSeekBg}{\celldown{14.29\%(1/7)}{27.17}} 
& \modelbg{DeepSeekBg}{\celldown{0.00\%(0/4)}{41.46}}  
& \modelbg{DeepSeekBg}{\cellup{100.00\%(3/3)}{58.54}} \\
& \modelbg{MiniMaxBg}{MiniMax-M2.5}  
& \modelbg{MiniMaxBg}{\shortstack[r]{72.60\%\\(SWE-bench-V)}}
& \modelbg{MiniMaxBg}{\val{52.00\%(39/75)}} 
& \modelbg{MiniMaxBg}{\celldown{29.27\%(12/41)}{22.73}} 
& \modelbg{MiniMaxBg}{\cellup{35.00\%(7/20)}{5.73}}  
& \modelbg{MiniMaxBg}{\celldown{14.29\%(1/7)}{14.98}} 
& \modelbg{MiniMaxBg}{\celldown{14.29\%(1/7)}{14.98}} 
& \modelbg{MiniMaxBg}{\celldown{0.00\%(0/4)}{29.27}}  
& \modelbg{MiniMaxBg}{\cellup{100.00\%(3/3)}{70.73}} \\
\midrule

\multirow{4}{*}{SWE-agent}
& \modelbg{GPTFourBg}{GPT-4o}        
& \modelbg{GPTFourBg}{\shortstack[r]{26.20\%\\(SEC-bench)}}
& \modelbg{GPTFourBg}{\val{21.33\%(16/75)}} 
& \modelbg{GPTFourBg}{\celldown{7.32\%(3/41)}{14.01}} 
& \modelbg{GPTFourBg}{\cellup{10.00\%(2/20)}{2.68}}  
& \modelbg{GPTFourBg}{\celldown{0.00\%(0/7)}{7.32}}  
& \modelbg{GPTFourBg}{\celldown{0.00\%(0/7)}{7.32}} 
& \modelbg{GPTFourBg}{\celldown{0.00\%(0/4)}{7.32}}  
& \modelbg{GPTFourBg}{\cellup{33.33\%(1/3)}{26.01}} \\
& \modelbg{GPTFiveBg}{GPT-5.2-Codex} 
& \modelbg{GPTFiveBg}{--}
& \modelbg{GPTFiveBg}{\val{49.33\%(37/75)}} 
& \modelbg{GPTFiveBg}{\celldown{46.34\%(19/41)}{2.99}}  
& \modelbg{GPTFiveBg}{\cellup{60.00\%(12/20)}{13.66}} 
& \modelbg{GPTFiveBg}{\celldown{28.57\%(2/7)}{17.77}} 
& \modelbg{GPTFiveBg}{\celldown{28.57\%(2/7)}{17.77}} 
& \modelbg{GPTFiveBg}{\celldown{25.00\%(1/4)}{21.34}} 
& \modelbg{GPTFiveBg}{\cellup{66.67\%(2/3)}{20.33}} \\
& \modelbg{DeepSeekBg}{DeepSeek-V4-Pro}    
& \modelbg{DeepSeekBg}{--}
& \modelbg{DeepSeekBg}{\val{54.67\%(41/75)}} 
& \modelbg{DeepSeekBg}{\celldown{48.78\%(20/41)}{5.89}}  
& \modelbg{DeepSeekBg}{\cellup{60.00\%(12/20)}{11.22}} 
& \modelbg{DeepSeekBg}{\celldown{28.57\%(2/7)}{20.21}} 
& \modelbg{DeepSeekBg}{\celldown{28.57\%(2/7)}{20.21}} 
& \modelbg{DeepSeekBg}{\celldown{25.00\%(1/4)}{23.78}} 
& \modelbg{DeepSeekBg}{\cellup{66.67\%(2/3)}{17.89}} \\
& \modelbg{MiniMaxBg}{MiniMax-M2.5}  
& \modelbg{MiniMaxBg}{--}
& \modelbg{MiniMaxBg}{\val{48.00\%(36/75)}} 
& \modelbg{MiniMaxBg}{\celldown{34.14\%(14/41)}{13.86}} 
& \modelbg{MiniMaxBg}{\cellup{40.00\%(8/20)}{5.86}}  
& \modelbg{MiniMaxBg}{\celldown{28.57\%(2/7)}{5.57}}  
& \modelbg{MiniMaxBg}{\celldown{14.29\%(1/7)}{19.85}} 
& \modelbg{MiniMaxBg}{\celldown{25.00\%(1/4)}{9.14}}  
& \modelbg{MiniMaxBg}{\cellup{100.00\%(3/3)}{65.86}} \\
\midrule

\multirow{4}{*}{Aider}
& \modelbg{GPTFourBg}{GPT-4o}        
& \modelbg{GPTFourBg}{\shortstack[r]{25.00\%\\(SWE-bench-L)}}
& \modelbg{GPTFourBg}{\val{21.33\%(16/75)}} 
& \modelbg{GPTFourBg}{\celldown{12.20\%(5/41)}{9.13}}  
& \modelbg{GPTFourBg}{\cellup{15.00\%(3/20)}{2.80}}  
& \modelbg{GPTFourBg}{\cellup{14.29\%(1/7)}{2.09}}  
& \modelbg{GPTFourBg}{\celldown{0.00\%(0/7)}{12.20}} 
& \modelbg{GPTFourBg}{\celldown{0.00\%(0/4)}{12.20}} 
& \modelbg{GPTFourBg}{\cellup{33.33\%(1/3)}{21.13}} \\
& \modelbg{GPTFiveBg}{GPT-5.2-Codex} 
& \modelbg{GPTFiveBg}{--}
& \modelbg{GPTFiveBg}{\val{45.33\%(34/75)}} 
& \modelbg{GPTFiveBg}{\celldown{26.83\%(11/41)}{18.50}} 
& \modelbg{GPTFiveBg}{\cellup{40.00\%(8/20)}{13.17}} 
& \modelbg{GPTFiveBg}{\celldown{0.00\%(0/7)}{26.83}} 
& \modelbg{GPTFiveBg}{\cellup{28.57\%(2/7)}{1.74}}  
& \modelbg{GPTFiveBg}{\celldown{0.00\%(0/4)}{26.83}} 
& \modelbg{GPTFiveBg}{\cellup{33.33\%(1/3)}{6.50}} \\
& \modelbg{DeepSeekBg}{DeepSeek-V4-Pro}    
& \modelbg{DeepSeekBg}{--}
& \modelbg{DeepSeekBg}{\val{28.00\%(21/75)}} 
& \modelbg{DeepSeekBg}{\celldown{9.76\%(4/41)}{18.24}}  
& \modelbg{DeepSeekBg}{\cellup{40.00\%(8/20)}{30.24}} 
& \modelbg{DeepSeekBg}{\celldown{0.00\%(0/7)}{9.76}}  
& \modelbg{DeepSeekBg}{\cellup{14.29\%(1/7)}{4.53}}  
& \modelbg{DeepSeekBg}{\celldown{0.00\%(0/4)}{9.76}}  
& \modelbg{DeepSeekBg}{\cellup{66.67\%(2/3)}{56.91}} \\
& \modelbg{MiniMaxBg}{MiniMax-M2.5}  
& \modelbg{MiniMaxBg}{--}
& \modelbg{MiniMaxBg}{\val{48.00\%(36/75)}} 
& \modelbg{MiniMaxBg}{\celldown{26.83\%(11/41)}{21.17}} 
& \modelbg{MiniMaxBg}{\celldown{15.00\%(3/20)}{11.83}} 
& \modelbg{MiniMaxBg}{\celldown{0.00\%(0/7)}{26.83}} 
& \modelbg{MiniMaxBg}{\celldown{0.00\%(0/7)}{26.83}} 
& \modelbg{MiniMaxBg}{\celldown{0.00\%(0/4)}{26.83}} 
& \modelbg{MiniMaxBg}{\cellup{33.33\%(1/3)}{6.50}} \\
\midrule

\multirow{4}{*}{AutoCodeRover}
& \modelbg{GPTFourBg}{GPT-4o}        
& \modelbg{GPTFourBg}{\shortstack[r]{38.40\%\\(SWE-bench-V)}}
& \modelbg{GPTFourBg}{\val{20.00\%(10/50)}} 
& \modelbg{GPTFourBg}{\celldown{14.29\%(5/35)}{5.71}}  
& \modelbg{GPTFourBg}{\cellup{20.00\%(4/20)}{5.71}}  
& \modelbg{GPTFourBg}{\celldown{0.00\%(0/6)}{14.29}} 
& \modelbg{GPTFourBg}{\celldown{0.00\%(0/4)}{14.29}} 
& \modelbg{GPTFourBg}{\celldown{0.00\%(0/2)}{14.29}} 
& \modelbg{GPTFourBg}{\cellup{33.33\%(1/3)}{19.04}} \\
& \modelbg{GPTFiveBg}{GPT-5.2-Codex} 
& \modelbg{GPTFiveBg}{--}
& \modelbg{GPTFiveBg}{\val{44.00\%(22/50)}} 
& \modelbg{GPTFiveBg}{\celldown{31.43\%(11/35)}{12.57}} 
& \modelbg{GPTFiveBg}{\cellup{45.00\%(9/20)}{13.57}} 
& \modelbg{GPTFiveBg}{\celldown{0.00\%(0/6)}{31.43}} 
& \modelbg{GPTFiveBg}{\celldown{0.00\%(0/4)}{31.43}} 
& \modelbg{GPTFiveBg}{\celldown{0.00\%(0/2)}{31.43}} 
& \modelbg{GPTFiveBg}{\cellup{66.67\%(2/3)}{35.24}} \\
& \modelbg{DeepSeekBg}{DeepSeek-V4-Pro}    
& \modelbg{DeepSeekBg}{--}
& \modelbg{DeepSeekBg}{\val{34.00\%(17/50)}} 
& \modelbg{DeepSeekBg}{\celldown{28.57\%(10/35)}{5.43}}  
& \modelbg{DeepSeekBg}{\cellup{35.00\%(7/20)}{6.43}}  
& \modelbg{DeepSeekBg}{\celldown{0.00\%(0/6)}{28.57}} 
& \modelbg{DeepSeekBg}{\celldown{0.00\%(0/4)}{28.57}} 
& \modelbg{DeepSeekBg}{\cellup{50.00\%(1/2)}{21.43}} 
& \modelbg{DeepSeekBg}{\cellup{66.67\%(2/3)}{38.10}} \\
& \modelbg{MiniMaxBg}{MiniMax-M2.5}  
& \modelbg{MiniMaxBg}{--}
& \modelbg{MiniMaxBg}{\val{26.00\%(13/50)}} 
& \modelbg{MiniMaxBg}{\celldown{20.00\%(7/35)}{6.00}}  
& \modelbg{MiniMaxBg}{\cellup{25.00\%(5/20)}{5.00}}  
& \modelbg{MiniMaxBg}{\celldown{16.67\%(1/6)}{3.33}} 
& \modelbg{MiniMaxBg}{\celldown{0.00\%(0/4)}{20.00}} 
& \modelbg{MiniMaxBg}{\celldown{0.00\%(0/2)}{20.00}} 
& \modelbg{MiniMaxBg}{\cellup{33.33\%(1/3)}{13.33}} \\
\midrule

\multirow{4}{*}{Agentless}
& \modelbg{GPTFourBg}{GPT-4o}        
& \modelbg{GPTFourBg}{\shortstack[r]{32.00\%\\(SWE-bench-L)}}
& \modelbg{GPTFourBg}{\val{18.00\%(9/50)}} 
& \modelbg{GPTFourBg}{\celldown{8.57\%(3/35)}{9.43}}   
& \modelbg{GPTFourBg}{\cellup{10.00\%(2/20)}{1.43}}  
& \modelbg{GPTFourBg}{\celldown{0.00\%(0/6)}{8.57}}  
& \modelbg{GPTFourBg}{\celldown{0.00\%(0/4)}{8.57}}  
& \modelbg{GPTFourBg}{\celldown{0.00\%(0/2)}{8.57}}  
& \modelbg{GPTFourBg}{\cellup{33.33\%(1/3)}{24.76}} \\
& \modelbg{GPTFiveBg}{GPT-5.2-Codex} 
& \modelbg{GPTFiveBg}{--}
& \modelbg{GPTFiveBg}{\val{24.00\%(12/50)}} 
& \modelbg{GPTFiveBg}{\celldown{22.86\%(8/35)}{1.14}}  
& \modelbg{GPTFiveBg}{\cellup{30.00\%(6/20)}{7.14}}  
& \modelbg{GPTFiveBg}{\celldown{0.00\%(0/6)}{22.86}} 
& \modelbg{GPTFiveBg}{\celldown{0.00\%(0/4)}{22.86}} 
& \modelbg{GPTFiveBg}{\celldown{0.00\%(0/2)}{22.86}} 
& \modelbg{GPTFiveBg}{\cellup{66.67\%(2/3)}{43.81}} \\
& \modelbg{DeepSeekBg}{DeepSeek-V4-Pro}    
& \modelbg{DeepSeekBg}{--}
& \modelbg{DeepSeekBg}{\val{40.00\%(20/50)}} 
& \modelbg{DeepSeekBg}{\celldown{28.57\%(10/35)}{11.43}} 
& \modelbg{DeepSeekBg}{\cellup{40.00\%(8/20)}{11.43}} 
& \modelbg{DeepSeekBg}{\celldown{16.67\%(1/6)}{11.90}}
& \modelbg{DeepSeekBg}{\celldown{0.00\%(0/4)}{28.57}}
& \modelbg{DeepSeekBg}{\celldown{0.00\%(0/2)}{28.57}} 
& \modelbg{DeepSeekBg}{\cellup{33.33\%(1/3)}{4.76}} \\
& \modelbg{MiniMaxBg}{MiniMax-M2.5}  
& \modelbg{MiniMaxBg}{--}
& \modelbg{MiniMaxBg}{\val{24.00\%(12/50)}} 
& \modelbg{MiniMaxBg}{\celldown{17.14\%(6/35)}{6.86}}
& \modelbg{MiniMaxBg}{\celldown{15.00\%(3/20)}{2.14}} 
& \modelbg{MiniMaxBg}{\celldown{16.67\%(1/6)}{0.47}} 
& \modelbg{MiniMaxBg}{\celldown{0.00\%(0/4)}{17.14}} 
& \modelbg{MiniMaxBg}{\celldown{0.00\%(0/2)}{17.14}} 
& \modelbg{MiniMaxBg}{\cellup{66.67\%(2/3)}{49.53}} \\
\bottomrule[1.2pt]
\end{tabular}
\end{adjustbox}

\end{table*}

\subsection{Experimental Setup}
\textbf{LLM/Agent.} We consider both LLM-based and Agent-based repairing methods from the leaderboard~\cite{swebench2024}. For LLM-based methods, Agentless~\cite{xia2024agentless} is included in the evaluation. For Agent-based methods, we cover SWE-agent~\cite{yang2024swe}, AutoCodeRover~\cite{zhang2024autocoderover}, OpenHands~\cite{wang2025openhands}, and Aider~\cite{aider}. Agentless and AutoCodeRover are evaluated only on Python projects because their current implementations support Python only.


\textbf{Models.}
We pair each applicable agent with four models:
GPT-4o\cite{gpt4o}, GPT-5.2-Codex\cite{gpt52codex}, DeepSeek-V4-Pro\cite{deepseekv4}, and MiniMax-M2.5\cite{minimaxm25}, resulting in 20 agent--model configurations.
All models are accessed through their official APIs.

\textbf{Execution Environment.}
Experiments are conducted on Ubuntu 24.04.3 LTS
(aarch64) with two ARM64 CPUs and 3.8\,GiB of memory, using
Conda v26.3.2 and Docker v29.3.1. 
More setup details are available on our project site.

\subsection{RQ2: Repair Correctness}
\label{subsec:repair_correctness}

\subsubsection{Effectiveness}

We evaluate repair correctness using Pass@1, a widely used metric in code-generation evaluation~\cite{chen2021evaluating}. In our setting, Pass@1 denotes the fraction of vulnerabilities for which an agent--model configuration produces a patch that passes both the functionality and security tests. 

\textbf{Results.} Table~\ref{tab:llm_loop_pass1} shows that LiL vulnerabilities are consistently harder to repair than LLM-ecosystem vulnerabilities and conventional vulnerabilities across different agents and backbone models.  For example, OpenHands with DeepSeek-V4-Pro drops from 60.00\% on LiL vulnerabilities to 41.46\% on LLM-ecosystem vulnerabilities, while OpenHands with MiniMax-M2.5 drops from 52.00\% to 29.27\%. These results suggest that the difficulty increase is not specific to a particular repair agent or model, but is a general challenge introduced by LiL vulnerabilities. 




The category-level results reveal substantial heterogeneity across LLM-mediated vulnerability mechanisms. C1 is relatively more tractable. For example, OpenHands with GPT-5.2-Codex achieves 75.00\% Pass@1 on C1. We conjecture the reason is that C1 vulnerabilities usually involve explicit and localized execution sinks, such as \texttt{exec}, \texttt{eval}, making the repair boundary easier for agents to identify. In contrast, C2--C4 are consistently harder, likely because their repair targets are less localized and require agents to reason about LLM-mediated trust-boundary propagation. For C2, the vulnerable behavior depends not only on a query execution sink, but also on how model-generated SQL, Cypher, or other query strings are constrained before reaching the database. For C3, correct repair requires controlling how LLM outputs are translated into privileged tool selections or tool arguments, which often involves preserving benign automation while blocking unsafe actions. C4 is especially challenging, receiving 0.00\% Pass@1 in 15 out of 20 situations. The reason is that unsafe model-output rendering requires agents to distinguish benign rich content from attacker-controlled HTML, Markdown, or UI payloads, making it difficult to harden the rendering boundary without breaking normal functionality. C5 obtains relatively high Pass@1 in several configurations, likely because output parsing or dispatch vulnerabilities often expose more explicit structured interfaces, such as callbacks, parsers, or schema-based handlers. These explicit interfaces can make the expected behavior easier to infer and constrain.


\subsubsection{Regression Analysis}

As multiple factors can affect the repair success, such as the agent architecture and backbone model, following previous studies~\cite{motwani2018automated}, we further conduct a regression analysis to examine whether the observed repair effectiveness remains after accounting for these factors, using the following equation.

\[
\operatorname{logit}\!\left(\Pr(Y_{ij}=1)\right)
=
\alpha+\beta S_i+\gamma_j+\boldsymbol{\delta}^{\top}\mathbf{X}_i ,
\]


where \(S_i\) indicates LLM-in-the-loop vulnerabilities, \(\gamma_j\) captures Agent--LLM fixed effects~(i.e., whether the the change of agent architecture can affect the conclusion), and \(\mathbf{X}_i\) denotes CVE-level controls. If the estimated Odds Ratio~(OR) for coefficient \(\beta\), \(\exp(\beta)\), is below 1, the vulnerabilities are less likely to be successfully repaired after controlling for these factors. Here, our considered CVE-level factors include programming language, severity and disclosure year.

\begin{table}[t]
\centering
\caption{Regression estimates for \dataset repair difficulty. CI denotes confidence interval. \textit{Forest plot} shows ORs with 95\% CIs. The dashed vertical line denotes OR \(=1\). }
\label{tab:repair_regression_forest}

\scriptsize
\setlength{\tabcolsep}{2pt}
\renewcommand{\arraystretch}{1.15}

\begin{tabularx}{\columnwidth}{
  @{}
  l
  >{\raggedright\arraybackslash}p{0.24\columnwidth}
  >{\raggedright\arraybackslash}X
  >{\centering\arraybackslash}p{0.30\columnwidth}
  r
  @{}
}
\toprule
\textbf{Model}
& \textbf{Controls}
& \textbf{OR (95\% CI)}
& \textbf{Forest plot }
& \textbf{\(p\)} \\
\midrule

M1 & None
& 0.578 (0.350, 0.957)
& \forestcell{0.578}{0.350}{0.957}
& 0.033 \\

M2 & Agent architecture
& 0.567 (0.325, 0.987)
& \forestcell{0.567}{0.325}{0.987}
& 0.045 \\

M3 & Language
& 0.586 (0.349, 0.983)
& \forestcell{0.586}{0.349}{0.983}
& 0.043 \\

M4 & CVSS
& 0.557 (0.335, 0.927)
& \forestcell{0.557}{0.335}{0.927}
& 0.024 \\

M5 & Year
& 0.595 (0.359, 0.986)
& \forestcell{0.595}{0.359}{0.986}
& 0.044 \\

\bottomrule
\end{tabularx}

\end{table}

\textbf{Results.} Table~\ref{tab:repair_regression_forest} summarizes the results. We can see that under different affect factors, the OR rates are consistently lower than 1~(with P-value less than 0.05, indicating the significance), which reflects that the low repair success of LiL vulnerabilities is not explained by differences in repair agents, backbone models, programming language, severity, or disclosure year. Moreover, we also explore the impact of factor combinations on the repair success rate, and find that the largest OR is 0.776~(combining all factors we considered), which is still less than 1, further confirming the repair difficulty is not affected by these factors.

\subsubsection{Patch Complexity}
\label{subsec:patch_complexity}

Motivated by the regression results, we examine whether LiL vulnerabilities require larger correct patches. For each ground-truth patch, we measure the number of changed lines, diff hunks, and modified files. Changed lines are calculated as the sum of added and deleted lines.

\begin{table}[t]
\centering
\caption{Ground-truth patch complexity (changed lines).}
\label{tab:patch_complexity_boundary}
\small
\setlength{\tabcolsep}{3.5pt}
\renewcommand{\arraystretch}{1.15}
\begin{tabular*}{\linewidth}{@{\extracolsep{\fill}}lrrrr}
\toprule
\textbf{Group} &
\makecell{\textbf{Median-}\\\textbf{lines}} &
\makecell{\textbf{Mean-}\\\textbf{lines}} &
\makecell{\textbf{Median-}\\\textbf{hunks}} &
\makecell{\textbf{Median-}\\\textbf{files}} \\
\midrule
\dataset & 43.0 & 56.4 & 3 & 1 \\
LLM-ecosystem & 14.0 & 28.8 & 2 & 1 \\
\bottomrule
\end{tabular*}
\end{table}

\textbf{Results.}
Table~\ref{tab:patch_complexity_boundary} shows that LiL vulnerabilities have substantially larger ground-truth patches. Their median and mean numbers of changed lines are 43.0 and 56.4, compared with 14.0 and 28.8 for LLM-ecosystem vulnerabilities. \dataset patches also contain slightly more hunks, while both groups modify a median of one file. Thus, the additional complexity mainly arises from a larger modification scope within the affected files rather than changes spanning more files. Furthermore, we analyze where does this additional complexity come from. To do so, we manually annotate all modified hunks in these patches and find that all of them fall entirely within LLM-specific repair boundaries. This suggests that the additional repair burden is associated with modifying security boundaries introduced by LLM-mediated execution.



\begin{tcolorbox}[size=title,opacityfill=0.1,breakable]
\noindent
\textbf{Answer to RQ2:} Existing methods can fix 25.8\%~(on average) of LiL vulnerabilities, which is 10.8\% and 15.5\% lower than fixing LLM-ecosystem and conventional software vulnerabilities, respectively.

\end{tcolorbox}

\subsection{RQ3: Repair Efficiency}

\begin{table}[h]
\centering
\caption{Average cost~(\$) per pass repair. \textit{Eco} indicates LLM-ecosystem.}
\label{tab:success-avg-cost-token}
\scriptsize
\renewcommand{\arraystretch}{1.08}
\setlength{\tabcolsep}{1.2pt}
\begin{adjustbox}{width=\columnwidth}
\begin{tabular}{l *{8}{>{\centering\arraybackslash}p{0.52cm}}}
\toprule
\multirow{2}{*}{\textbf{Agent}} &
\multicolumn{2}{c}{\textbf{GPT-4o}} &
\multicolumn{2}{c}{\textbf{GPT-5.2-C}} &
\multicolumn{2}{c}{\textbf{DS-V4}} &
\multicolumn{2}{c}{\textbf{MM-M2.5}} \\
\cmidrule(lr){2-3}\cmidrule(lr){4-5}\cmidrule(lr){6-7}\cmidrule(lr){8-9}
 & \textbf{Eco.} & \textbf{LiL}
 & \textbf{Eco.} & \textbf{LiL}
 & \textbf{Eco.} & \textbf{LiL}
 & \textbf{Eco.} & \textbf{LiL} \\
\midrule
OpenHands &
\cellcolor{CostPeach!31}4.12 & \cellcolor{CostPeach!60}10.11 &
\cellcolor{CostPeach!15}0.83 & \cellcolor{CostPeach!14}0.51 &
\cellcolor{CostPeach!12}0.05 & \cellcolor{CostPeach!12}0.07 &
\cellcolor{CostPeach!12}0.09 & \cellcolor{CostPeach!12}0.21 \\
SWE-Agent &
\cellcolor{CostPeach!25}2.98 & \cellcolor{CostPeach!59}10.00 &
\cellcolor{CostPeach!16}1.06 & \cellcolor{CostPeach!15}0.76 &
\cellcolor{CostPeach!12}0.07 & \cellcolor{CostPeach!12}0.09 &
\cellcolor{CostPeach!12}0.06 & \cellcolor{CostPeach!12}0.09 \\
Aider &
\cellcolor{CostPeach!20}1.76 & \cellcolor{CostPeach!20}1.77 &
\cellcolor{CostPeach!14}0.59 & \cellcolor{CostPeach!15}0.71 &
\cellcolor{CostPeach!12}0.05 & \cellcolor{CostPeach!12}0.09 &
\cellcolor{CostPeach!12}0.06 & \cellcolor{CostPeach!12}0.12 \\
AutoCodeRover &
\cellcolor{CostPeach!19}1.53 & \cellcolor{CostPeach!16}1.04 &
\cellcolor{CostPeach!16}1.04 & \cellcolor{CostPeach!16}0.99 &
\cellcolor{CostPeach!12}0.12 & \cellcolor{CostPeach!12}0.08 &
\cellcolor{CostPeach!12}0.26 & \cellcolor{CostPeach!13}0.28 \\
Agentless &
\cellcolor{CostPeach!25}2.87 & \cellcolor{CostPeach!25}2.86 &
\cellcolor{CostPeach!22}2.32 & \cellcolor{CostPeach!23}2.51 &
\cellcolor{CostPeach!12}0.09 & \cellcolor{CostPeach!12}0.13 &
\cellcolor{CostPeach!12}0.13 & \cellcolor{CostPeach!12}0.19 \\
\bottomrule
\end{tabular}
\end{adjustbox}
\end{table}

\begin{figure}[h]
\centering
\includegraphics[width=0.48\textwidth]{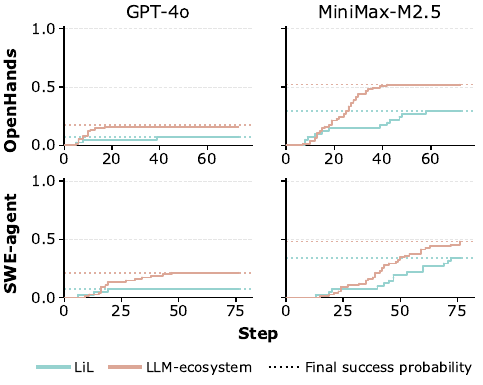}
\caption{Cumulative repair success by step. Due to space limitation, we report four representative results, others exhibit similar patterns. Full results are available on our project site.}
\label{fig:step}
\end{figure}

\textbf{Results.}
Table~\ref{tab:success-avg-cost-token} presents the cost of patching \dataset vulnerabilities, and Figure~\ref{fig:step} depicts the steps needed for generating correct patches. Overall, the results demonstrate that repairing LiL vulnerabilities requires more steps compared to repairing LLM-ecosystem vulnerabilities. Regarding the token cost~(\$), it is difficult to judge which type of vulnerability requires is more costly as the results are fluctuating. However, regarding the steps needed for generating correct patches, the results showcase that across most configurations, LLM-ecosystem vulnerability repairs accumulate successes faster and reach higher final success probabilities than LiL vulnerability repairs. For instance, under OpenHands with DeepSeek-V4-Pro, the adjacent setting reaches about 0.60 final success probability, while the loop setting reaches about 0.42; under OpenHands with MiniMax-M2.5, the gap is about 0.52 versus 0.29. These results suggest that LLM-in-the-loop vulnerabilities are not only harder to repair in terms of final Pass@1, but also tend to require more repair effort and slower successful convergence.


\begin{tcolorbox}[size=title,opacityfill=0.1,breakable]
\noindent
\textbf{Answer to RQ3:} Repairing a LiL vulnerability costs \$0.07-\$10.11, showing no significant difference compared to fixing other types of vulnerabilities. However, under the same step budget, agents consistently achieve fewer successful LiL repairs, indicating lower repair efficiency.

\end{tcolorbox}

\subsection{RQ4: Failure Analysis}


\begin{table*}[t]
\centering
\caption{Failure causes on LiL vulnerabilities.  Results are combined across all models.}
\label{tab:strict_failure_causes}
\small
\setlength{\tabcolsep}{4pt}
\renewcommand{\arraystretch}{1.08}
\resizebox{\textwidth}{!}{
\begin{tabular}{lrrrrrr}
\toprule
\textbf{Agent} &
\multicolumn{1}{c}{\textbf{Generation}} &
\multicolumn{5}{c}{\textbf{Validation}} \\
\cmidrule(lr){2-2}
\cmidrule(lr){3-7}
&
\textbf{No Patch} &
\textbf{Build Error} &
\textbf{Wrong Location} &
\textbf{Security Only} &
\textbf{Functionality Only} &
\textbf{Both Test Fail} \\
\midrule
Agentless &
13.3\% (15/113) &
9.7\% (11/113) &
15.0\% (17/113) &
\cellcolor{FailurePurple!37}37.2\% (42/113) &
24.8\% (28/113) &
15.0\% (17/113) \\

Aider &
29.3\% (39/133) &
14.3\% (19/133) &
20.3\% (27/133) &
\cellcolor{FailurePurple!43}42.9\% (57/133) &
11.3\% (15/133) &
4.5\% (6/133) \\

ACR &
26.2\% (28/107) &
11.2\% (12/107) &
9.3\% (10/107) &
\cellcolor{FailurePurple!27}27.1\% (29/107) &
24.3\% (26/107) &
11.2\% (12/107) \\

OpenHands &
9.1\% (10/110) &
20.9\% (23/110) &
13.6\% (15/110) &
\cellcolor{FailurePurple!39}39.1\% (43/110) &
23.6\% (26/110) &
12.7\% (14/110) \\

SWE-agent &
9.3\% (10/108) &
20.4\% (22/108) &
17.6\% (19/108) &
\cellcolor{FailurePurple!44}44.4\% (48/108) &
23.1\% (25/108) &
3.7\% (4/108) \\
\bottomrule
\end{tabular}
}
\end{table*}

To understand why agents fail on LiL vulnerabilities, we analyze all unsuccessful repair attempts. Based on our definition in Section~\ref{sec:definition-of-llmloop} and following prior studies on vulnerability repair and patch correctness analysis~\cite{bui2022vul4j, yu2019alleviating, qi2015analysis},  we classify failures based on the generated patches and evaluation logs into one generation-stage category and five validation-stage categories. \textit{No Patch} means that the agent fails to produce a usable candidate patch within the given budget. \textit{Build Error} means that the generated patch cannot be successfully applied to the repository or fails to build. \textit{Wrong Location} means that the patch modifies an irrelevant or incorrect code location. \textit{Security Only} denotes patches that pass regression tests but still fail security tests. \textit{Functionality Only} denotes patches that fix the security test but break benign functionality. \textit{Both Test Fail} denotes patches that fail both the security tests and the regression tests.


\textbf{Results.} \textbf{Data Analysis.}
Table~\ref{tab:strict_failure_causes} shows that LiL repair is primarily a validation-stage problem. The dominant failure mode is security-only failure, ranging from 27.1\% to 44.4\% of failed cases, indicating that agents frequently generate patches that still leave part of the LLM-mediated exploit path exploitable. Functionality-only failures are also notable, accounting for 11.3\%--24.8\% of failed cases, suggesting that some patches block unsafe behavior only by damaging benign functionality. In contrast, simultaneous security and regression failures are relatively uncommon, mostly below 15.0\%. This pattern suggests that agents usually do not fail by producing entirely broken patches; rather, they often optimize for only one side of the repair objective. The key challenge is therefore to balance two competing requirements: enforcing a security boundary over model-mediated behavior while preserving the intended LLM-enabled functionality.

\textbf{Case Analysis.} To explain the failure patterns quantitatively observed above, we select
two representative failed repairs from C2 and C4, the categories with the
lowest Pass@1. 

\begin{figure}[t]
\centering

\begin{subfigure}[t]{0.48\textwidth}
    \centering
    \includegraphics[width=\linewidth]{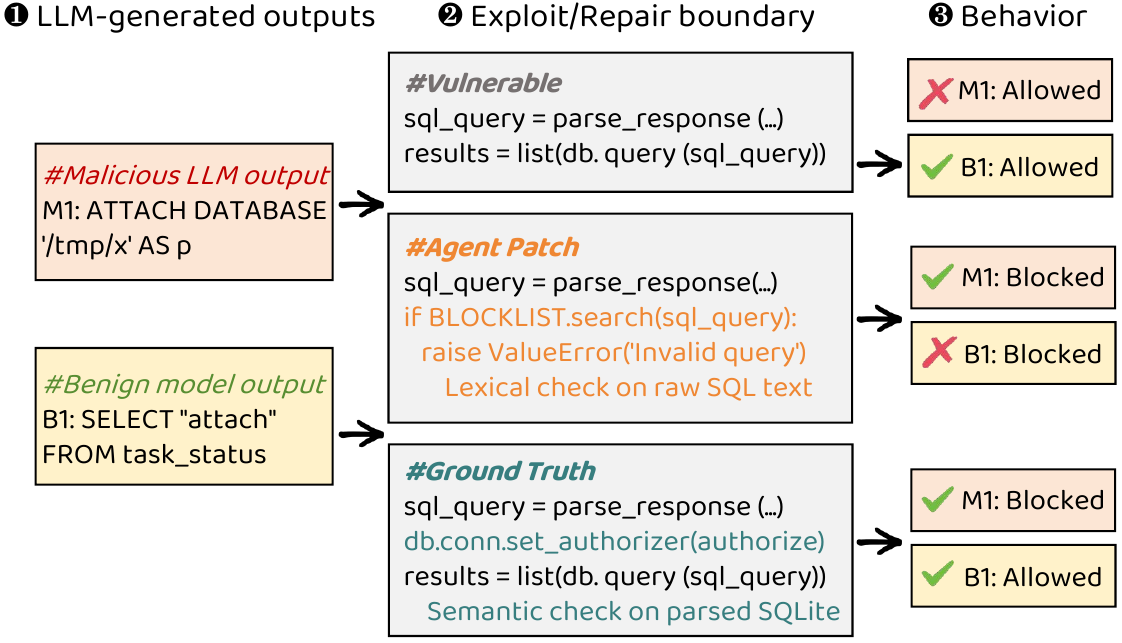}
    \caption{CVE-2024-12911}
    \label{fig:case1}
\end{subfigure}
\hfill
\begin{subfigure}[t]{0.48\textwidth}
    \centering
    \includegraphics[width=\linewidth]{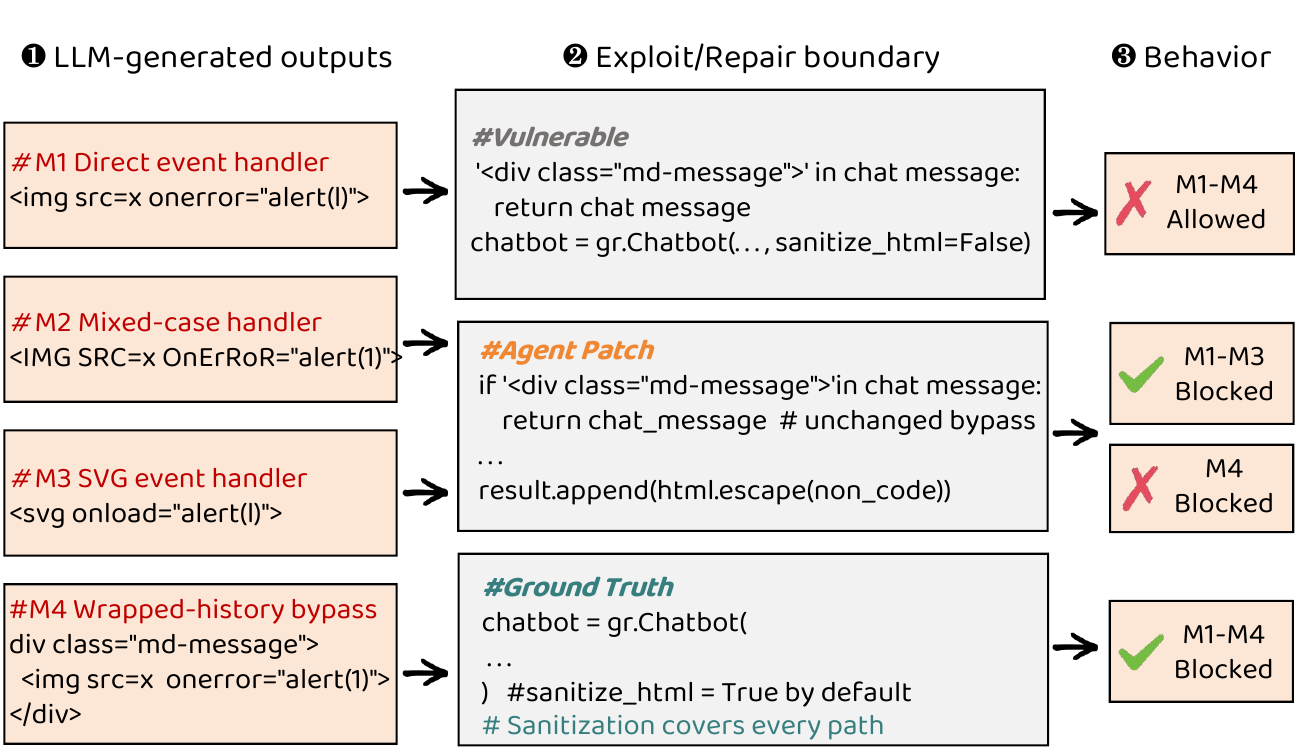}
    \caption{CVE-2024-3402}
    \label{fig:case2}
\end{subfigure}

\caption{Case studies of representative vulnerabilities.}
\label{fig:case_studies}
\end{figure}



\textbf{Case1--C2.} Figure~\ref{fig:case1} illustrates CVE-2024-12911 in LlamaIndex, where LLM-generated SQL is directly executed by SQLite, allowing a malicious \texttt{ATTACH DATABASE} query to create arbitrary files. The agent blocks dangerous keywords using a regular expression, but this lexical check also rejects benign queries containing an identifier such as \texttt{"attach"}. In contrast, the developer patch uses SQLite's authorizer to deny actual unsafe actions such as \texttt{SQLITE\_ATTACH} while preserving legitimate reads. This case shows that correct repair requires semantic enforcement at the downstream execution boundary rather than keyword filtering over model output.

\textbf{Case2--C4.} Figure~\ref{fig:case2} illustrates CVE-2024-3402, a stored XSS vulnerability in ChuanhuChatGPT, where LLM-generated HTML reaches a Gradio chatbot configured with sanitization disabled. The agent escapes HTML inside the normal message-conversion loop, blocking three attack variants, but an already-wrapped message triggers an unchanged early return and bypasses the new defense. Consequently, one security test remains exploitable. In contrast, the ground-truth patch restores sanitization at the final rendering sink, ensuring that all message paths are sanitized before display.

These cases show that agents often treat LiL vulnerabilities as local output filtering problems. Their patches operate on the surface form of LLM-controlled content, such as keywords in generated SQL or HTML in a specific message path. However, the actual security risk appears when this content crosses a downstream semantic boundary, such as database execution or UI rendering. As a result, repairs at the output surface may block visible attack variants but still miss equivalent behaviors at the sink, or they may reject benign cases that share similar lexical forms. This explains why many generated patches satisfy only one validation objective. The main repair challenge is therefore boundary completeness rather than simple vulnerability recognition.

\begin{tcolorbox}[size=title,opacityfill=0.1,breakable]
\noindent
\textbf{Answer to RQ4:} On average, 72.3\% of the generated patches can only pass security or functional testing. Multi-objective satisfaction is a key challenge in fixing LiL vulnerabilities.

\end{tcolorbox}

\section{Discussion}

\subsection{Implication}

Based on our analysis, we provide implications for different stakeholders in the large language model software ecosystem.

\textit{Implications for vulnerability benchmarks. }
Our results show that most vulnerabilities collected from LLM ecosystem projects do not necessarily involve behavior unique to LLMs. As a result, existing dataset based studies on LLM security mainly provide a traditional software security view of vulnerabilities in the LLM ecosystem, rather than capturing the distinct risks introduced when LLMs participate in the vulnerable execution chain. This suggests that future LLM security benchmarks should not treat all vulnerabilities in LLM-related repositories as a homogeneous class. Instead, benchmarks should explicitly distinguish whether the LLM participates in the vulnerable execution chain, and annotate how model generated or model mediated content reaches downstream security sensitive logic.

\textit{Implications for automated repair agents.}
The repair results indicate that LiL vulnerabilities tend to be harder and less efficient for agents to fix. Therefore, improving repair performance requires more than stronger backbone models. Agents need mechanisms for identifying trust boundaries mediated by LLMs, reasoning about how content generated by models propagates to downstream sinks, and validating both security and benign functionality enabled by LLMs.

\textit{Implications for LLM application development.}
Failure analysis reveals a core repair trade-off in which agents often preserve benign functionality while leaving the exploit path open, or block the exploit by breaking intended LLM enabled behavior. This highlights the need for developers to make LLM boundaries explicit and testable. Security tests should also cover both malicious model behaviors and benign functionality, so that repairs can be evaluated against the full trust boundary rather than a single exploit instance.


\subsection{Threat to Validity}

\textit{Internal Validity.} Internal threats arise from the manual identification of LLM-in-the-loop vulnerabilities and patch-hunk annotation, which may introduce bias. Refer to previous research~\cite{shi2022large,croft2023data}, We mitigate this through independent analysis by multiple researchers and clearly defined annotation criteria, reducing the potential impact on our results.

\textit{Construct Validity.} We do not invoke live LLMs during patch validation. Instead, we use predefined malicious output variants derived from the original PoCs. Although these variants cannot exhaust all possible LLM outputs, they eliminate sampling uncertainty and enable consistent, reproducible evaluation. Our study focuses on whether a patch repairs the downstream vulnerability once malicious model output has been produced. Whether a live model can be induced to generate malicious outputs through jailbreak or prompt injection is an upstream security issue beyond our scope.

\textit{External Validity.} Our considered LLM-relevant components mainly come from Huntr, which may ignore some components from other sources. Our rationale is twofold. First, Huntr covers diverse components, from deep learning frameworks to more basic computing frameworks, enabling an in-depth analysis of LLM-relevant components. Second, as a bug bounty platform, Huntr contains components with practical security relevance. Since \dataset is the first dataset designed to support the study of LLM-in-the-loop vulnerabilities, we believe it can facilitate future research in this domain.


\section{Related Work}

\subsection{LLM Vulnerability Analysis}

To better understand software vulnerabilities, extensive research has been conducted in both traditional software systems and LLM-based applications~\cite{zhou2025large}. For traditional software vulnerabilities, prior work has extensively studied vulnerability characteristics~\cite{shahzad2012large, bhuiyan2021security, gonzalez2019automated} and constructed repair benchmarks~\cite{bui2022vul4j, just2014defects4j, hu2025sokVul4C, wu2023effective} to evaluate the difficulty of fixing vulnerabilities and the effectiveness of automated repair techniques. For LLM-based applications, existing studies analyze vulnerabilities from multiple perspectives, including LLM supply chains~\cite{wang2025large, ma2025understanding, wang2025sok, hu2025understanding, huang2024lifting, luna2026security, moia2025analysis}, LLM applications~\cite{zhao2025llm, yan2024exploring, su2025gpt, xie2025llm, hou2025security, antebi2024gpt, iqbal2024llm, tao2023opening, shao2024llms, hui2024pleak, liu2023prompt, pedro2023prompt}, the LLM itself~\cite{das2025security, abdali2024securing, chitimoju2024survey, andriushchenko2025jailbreaking, dong2025sata}, and individual components (e.g., LLM agent frameworks)~\cite{gasmi2026bridging, he2025llm, zhang2025agent}. Model-centric studies focus on robustness issues such as jailbreak, adversarial attacks, and prompt injection, while system-level analyses examine risks across the LLM supply chain and its complex dependencies.

However, these studies typically analyze vulnerabilities in an aggregated manner without distinguishing whether they arise from traditional software logic or from LLM involvement (i.e., LLM-in-the-loop vulnerabilities). Moreover, they primarily focus on vulnerability categories and distributions, with limited attention to repairability. In contrast, our work explicitly focuses on LLM-in-the-loop vulnerabilities by providing a clear definition, constructing the first dedicated benchmark, and systematically analyzing both their characteristics and the repair performance of SOTA LLM-based agents.

\subsection{LLM-based Vulnerability Repair}

Recent advances in LLMs have demonstrated promising capabilities in Automated Program Repair (APR). Early studies primarily explored zero-shot repair and prompt engineering, showing that while LLMs can generate plausible patches, their effectiveness remains limited for real-world vulnerabilities~\cite{pearce2023examining, noever2023can}. To improve repair accuracy, subsequent work incorporates context optimization, model fine-tuning, and program analysis techniques. These approaches enhance the ability of LLMs to handle complex code semantics and localized vulnerability patterns, but still struggle with broader program understanding and scalability~\cite{zhang2024evaluating, de2024enhanced, berabi2024deepcode}. More recently, research has shifted toward agent-based and multi-step repair frameworks, where LLMs iteratively perform fault localization, patch generation, and validation. Such approaches significantly improve repair performance on traditional software vulnerabilities~\cite{fakih2025llm4cve, nong2025appatch, kim2025logs, yu2025patchagent}. However, these works primarily focus on repairing vulnerabilities in traditional software systems, whereas our work targets LLM-in-the-loop vulnerabilities that arise from the interaction between LLMs and surrounding components.

\section{Conclusion}

In this work, we have defined two types of vulnerabilities, LiL vulnerability and LLM-ecosystem vulnerability, and presented \dataset, a benchmark dataset containing 116 of these vulnerabilities. Using \dataset, we have conducted a comprehensive study to assess the capabilities of existing LLM-based methods in repairing vulnerabilities in \dataset. Moreover, we have identified core reasons for agent repair failures and derived insightful implications for different stakeholders. We hope this research and our dataset can facilitate future research on boosting the security of LLM-integrated systems.


\bibliographystyle{IEEEtran}
\bibliography{refs}

\end{document}